\documentclass[a4paper,11pt]{article}
\pdfoutput=1
\usepackage{jheppub}

\usepackage{amssymb,amsmath,amsfonts}
\usepackage[normalem]{ulem}
\usepackage[utf8x]{inputenc}
\usepackage{slashed}
\usepackage{graphicx}
\usepackage{tabularx}
\usepackage{here}
\usepackage{color}
\usepackage{csquotes} 
\usepackage{comment}
\usepackage{mathrsfs}
\usepackage{float}
\usepackage{ascmac}
\usepackage{multirow}
\usepackage{longtable}
\usepackage{bm}
\usepackage{ulem}

\usepackage[italicdiff]{physics}

\usepackage[hang,small,bf]{caption}
\usepackage[subrefformat=parens]{subcaption}
\makeatletter
\newcommand*\rel@kern[1]{\kern#1\dimexpr\macc@kerna}
\newcommand*\widebar[1]{%
  \begingroup
  \def\mathaccent##1##2{%
    \rel@kern{0.8}%
    \overline{\rel@kern{-0.8}\macc@nucleus\rel@kern{0.2}}%
    \rel@kern{-0.2}%
  }%
  \macc@depth\@ne
  \let\math@bgroup\@empty \let\math@egroup\macc@set@skewchar
  \mathsurround\z@ \frozen@everymath{\mathgroup\macc@group\relax}%
  \macc@set@skewchar\relax
  \let\mathaccentV\macc@nested@a
  \macc@nested@a\relax111{#1}%
  \endgroup
}
\makeatother

\numberwithin{equation}{section}

\preprint{
\begin{minipage}{5cm}
\small
\flushright
EPHOU-25-019\\
KYUSHU-HET-345
\end{minipage}}

\title{Generalized CP from non-invertible selection rules}

\author{Tatsuo Kobayashi$^{1}$ and} 
\author{Hajime Otsuka$^{2,3}$}
\affiliation{
$^1$Department of Physics, Hokkaido University, Sapporo 060-0810, Japan}
\affiliation{
$^2$Department of Physics, Kyushu University, 744 Motooka, Nishi-ku, Fukuoka 819-0395, Japan}
\affiliation{
$^3$Quantum and Spacetime Research Institute (QuaSR), Kyushu University, 744 Motooka, Nishi-ku, Fukuoka 819-0395, Japan}
\emailAdd{kobayashi@particle.sci.hokudai.ac.jp}
\emailAdd{otsuka.hajime@phys.kyushu-u.ac.jp}

\abstract{
We study a framework in which fields are labeled by basis elements of a fusion algebra with non-invertible fusion rules. 
In particular, we consider the case where fields are labeled by conjugacy classes of a finite group rather than its irreducible representations. 
When the fusion rules possess a $\mathbb{Z}_2$ symmetry identified with charge conjugation, a CP-invariant system can be consistently defined together with parity transformation. 
Furthermore, it is found that combining group-based flavor symmetries underlying non-invertible selection rules with CP symmetry naturally leads to a generalized CP transformation. 
We also demonstrate the possibility of spontaneous CP violation in this framework and discuss its implications for Yukawa textures. 
}

\makeatletter
\gdef\@fpheader{}
\makeatother

\begin{document}

\maketitle


\section{Introduction}

The flavor structure of quarks and leptons remains unexplained in the Standard Model of particle physics. 
So far, flavor symmetries, in particular discrete flavor symmetries, have been extensively studied to address the peculiar pattern of fermion masses and mixings (see, e.g., Refs.~\cite{Altarelli:2010gt,Ishimori:2010au,Hernandez:2012ra,King:2013eh,King:2014nza,Tanimoto:2015nfa,King:2017guk,Petcov:2017ggy,Feruglio:2019ybq,Kobayashi:2022moq} for reviews). 
Furthermore, some of the discrete flavor symmetries can be compatible with the generalized CP symmetry~\cite{Holthausen:2012dk,Chen:2014tpa}, leading to an enlarged symmetry group of the form $G_f \rtimes \mathbb{Z}_2^{\mathrm{CP}}$~\cite{Feruglio:2012cw}. 
Since the model parameters are restricted to take real numbers, this framework yields testable predictions (see, e.g., Refs.~\cite{Holthausen:2012dk,Ding:2013bpa}, for the $A_4$ case). 

Recently, an alternative approach has been explored to understand the origin of flavor structure of quarks and leptons. 
In the conventional approach based on group-based symmetries, the fields are assumed to transform as some irreducible representations of a given group. However, it has been recently proposed that fields are labeled by basis elements of a fusion algebra endowed with an associative multiplication law: 
\begin{align}
    ab = \sum_{c\in A} N_{ab}^{c}c.
\end{align}
Here, $A=\{e, a, b,...\}$ represents a finite set of basis elements with $e$ being the unit, and the coefficients $N_{ab}^{c}$ are taken to be non-negative integers. 
In the context of field theory and string theory, this setup will correspond to the situation in which fields are labeled by conjugacy classes of a group rather than by its representations. Since fusion rules generically yield multiple terms in the product of two basis elements, the associated selection rules for fields labeled by these elements exhibit a non-invertible structure. 
For instance, a scattering process involving fields $\phi_a$ and $\phi_b$, labeled by the elements $a$ and $b$, is governed by the coefficient $N_{ab}^c$ in the fusion algebra. 
These non-invertible selection rules lead to a structure that is richer than that of ordinary group-based symmetries.

It was known that these fusion rules are compatible with field theory and string theory (see, e.g., Ref.~\cite{Kaidi:2024wio}). 
For instance, non-invertible selection rules can be realized on toroidal orbifolds in heterotic string theory~\cite{Kobayashi:2025ocp} and type II D-brane models~\cite{Kobayashi:2024yqq,Funakoshi:2024uvy}, where topological operators generate non-invertible symmetries together with the corresponding non-invertible selection rules. 
In the context of field theory, the non-invertible structure plays a significant role in particle phenomenology such as the realization of tiny neutrino mass~\cite{Cordova:2022fhg,Kobayashi:2025cwx}, and new Yukawa textures for quarks~\cite{Kobayashi:2025znw} and leptons~\cite{Kobayashi:2025ldi}. 
These non-invertible selection rules arising from fusion algebras may contain group-like symmetries, such as $\mathbb{Z}_N$, which may remain stable under quantum corrections. The existence of such group-like symmetries has been demonstrated for fusion rules associated with the conjugacy classes of various groups~\cite{Dong:2025jra}, and further supported through spurion analysis~\cite{Suzuki:2025bxg,Suzuki:2025kxz}. 
It naturally raises the question of whether a $\mathbb{Z}_2$ symmetry appearing in the non-invertible selection rules can be identified with charge conjugation. 
However, the relation between the CP transformation and non-invertible selection rules has not been fully clarified yet. 
This motivates us to construct a CP-invariant framework in which fields are labeled by specific basis elements of a fusion algebra that admits a $\mathbb{Z}_2$ symmetry.

In this paper, we identify a $\mathbb{Z}_2$ symmetry underlying certain non-invertible selection rules with the charge conjugation, where the charge conjugate of a field $\phi$ is likewise labeled by an element of the fusion algebra. 
By combining the charge conjugation with the parity transformation, one can realize the four-dimensional CP transformation. 
Furthermore, some fusion algebras admit additional group-based symmetries $G_f$. 
When fields are labeled to irreducible representations of $G_f$ in a manner consistent with CP, the full 
symmetry of this system is described by $G_f\rtimes \mathbb{Z}_2^{\mathrm{CP}}$. Hence, the generalized CP symmetry can be formulated in theories whose non-invertible selection rules are governed by fusion algebras.

This paper is organized as follows. 
In Sec.~\ref{sec:2}, we formulate the generalized CP for fusion rules of conjugacy classes of a group which admits the group-based flavor symmetries and the charge conjugation. It leads to the condition for couplings of fields consistent with the generalized CP. 
In Sec.~\ref{sec:ex}, we explicitly demonstrate the generalized CP for concrete conjugacy classes of groups. 
When we combine two fusion algebras, one can realize the texture zero of Yukawa couplings in the context of generalized CP, as discussed in Sec.~\ref{sec:textures}. 
Furthermore, the spontaneous CP violation is explicitly demonstrated in Sec.~\ref{sec:spontaneous}. 
We demonstrate that the three-zero textures still hold even after the spontaneous CP violation. 
Finally, Sec.~\ref{sec:con} is devoted to the conclusions.

\section{Formulations}
\label{sec:2}

We formulate a generalized CP transformation underlying non-invertible selection rule in Sec.~\ref{sec:gCP} and its implications for couplings in Sec.~\ref{sec:couplings}

\subsection{Generalized CP symmetry}
\label{sec:gCP}

Let us suppose that a field in four-dimensional quantum field theory is labeled by a conjugacy class $[g]$ of a finite group $G$ rather than its representations. The conjugate $\phi^\ast$ is then labeled by a class $\overline{[g]}=[g^{-1}]$. 
Here and in what follows, we omit Lorentz indices unless explicitly specified. 
When the fields $\phi_i$ are labeled by the conjugacy class $[g_i]$, their selection rules are governed by a hypergroup. 
In particular, the conjugacy classes form a commutative fusion algebra with fusion rules:
\begin{align}
\label{eq:rules}
    [g_i]\cdot[g_j]=\sum_k N_{ij}^k [g_k],
\end{align}
with $N_{ij}^k$ being non-negative integers. 
Note that when we take $[g_j]$ as $[g_i^{-1}]$ in the left-handed side, the identity class $[e]$ appears on the right-handed side, i.e., $[g_i]\cdot [g_i^{-1}]=[e]+\dots$.\footnote{Here, we normalize the conjugacy classes such that $N_{[g_i],[g_i^{-1}]}^{[e]}=1$.}
With this structure, the bare interaction $\phi_1\phi_2\cdots \phi_n$ is allowed if and only if there exist representatives $\tilde{g}_i\in [g_i]$ whose product contains the identity element $e\in G$:
\begin{align}
    \tilde{g}_1\tilde{g}_2\cdots\tilde{g}_n = e.
\end{align}

It was known that the selection rules associated with the conjugacy class are often non-invertible, while nevertheless include group-like symmetries at tree and/or loop-levels~\cite{Heckman:2024obe,Kaidi:2024wio,Kobayashi:2024yqq}. 
As an illustrative example, let us suppose that the selection rules enjoy the following symmetry based on a finite group $G_f$, 
under which the conjugacy classes transform as
\begin{align}
[g_i] \rightarrow \rho(g)_{ij} [g_j],
\label{eq:flavor}
\end{align}
where $\rho(g)$ denotes a representation matrix of $G_f$ for an element $g\in G_f$. 
It means that the fusion algebra \eqref{eq:rules} is invariant under $G_f$ provided that
\begin{align}
    \rho(g)_{i\ell}\rho(g)_{jm}N_{lm}^k = N_{ij}^k
\end{align}
holds for all $g\in G_f$. 
In this case, a four-dimensional field $\phi$ labeled by $[g_i]$ transforms in the same way as the corresponding class:
\begin{align}
    \phi_i \xrightarrow{g} \rho(g)_{ij} \phi_j.
\end{align}
It is nothing but a flavor symmetry acting on the fields which belong to a certain irreducible representation of the group $G_f$. 
Note that this flavor symmetry does not arise from an ordinary invertible symmetry of the theory, but instead emerges from the underlying non-invertible selection rules described by the fusion algebra of conjugacy classes.

In addition to the flavor transformations of the conjugacy classes~\eqref{eq:flavor}, other types of discrete symmetries may also exist. 
In particular, let us discuss the following $\mathbb{Z}_2$ symmetry:\footnote{We explicitly demonstrate such a symmetry for specific selection rules in Sec.~\ref{sec:ex}.}
\begin{align}
    [g_i] \rightarrow X^{\mathrm{CP}}_{ij}[g_j^{-1}], 
\label{eq:CPclass}
\end{align}
where $X^{\mathrm{CP}}$ is a unitary matrix. 
Since the field $\phi_i$ and the conjugate $\bar{\phi}_i$ are respectively assigned to $[g_i]$ and $[g_i^{-1}]$, the transformation~\eqref{eq:CPclass} can be identified with the charge conjugation in four-dimensional quantum field theories. 
If $[g_i^{-1}]$ coincides with $[g_i]$ in a basis $X^{\mathrm{CP}}=\mathbb{I}$, the corresponding fields are self conjugate under CP, 
as appeared in the fusion algebras associated with $\mathbb{Z}_2$ gauging of $\mathbb{Z}_N$ symmetry~\cite{Kobayashi:2024yqq,Kobayashi:2024cvp}. 
Hence, by combining the charge conjugation with the parity transformation, we would identify the underlying $\mathbb{Z}_2$ symmetry of the non-invertible selection rules with CP symmetry. At the level of four-dimensional fields, it acts as
\begin{align}
    \phi(x) \rightarrow  X^{\mathrm{CP}}\phi^\ast (x_P),
\end{align}
with $x=(t,\mathbf{x})$ and $x_P=(t,-\mathbf{x})$. 
Note that $X^{\mathrm{CP}}$ is a unitary matrix, and the canonical CP transformation corresponds to $X^{\mathrm{CP}}=\mathbb{I}$. 
While the above expression describes the CP transformation of scalar fields, an additional matrix acting on spinor indices must be included in the case of fermions.

It was shown in Refs.~\cite{Feruglio:2012cw,Holthausen:2012dk} that the unitary matrix $X^{\mathrm{CP}}$ satisfies the consistency condition. 
By performing the following chain:
\begin{align}
    \phi(x) \xrightarrow{\mathrm{CP}} X^{\mathrm{CP}}\bar{\phi}(x_P) \xrightarrow{g} X^{\mathrm{CP}}\rho^{\ast}(g)\bar{\phi}(x_P) \xrightarrow{\mathrm{CP}^{-1}} X^{\mathrm{CP}}\rho^{\ast}(g)(X^{\mathrm{CP}})^{-1}\phi(x),
\end{align}
we find 
\begin{align}
    X^{\mathrm{CP}}\rho^{\ast}(g)(X^{\mathrm{CP}})^{-1} = \rho(g'),
\end{align}
with $g,g'\in G$. 
The above consistency condition should be simultaneously satisfied for all irreducible representations of the flavor group. 
When $X^{\mathrm{CP}}$ is a symmetric matrix, the flavor group and CP transformation are enlarged to $G_f\rtimes \mathbb{Z}_2^{\mathrm{CP}}$ generated by the generalized CP transformation.

\subsection{CP invariance for couplings}
\label{sec:couplings}

We examine whether the CP transformation \eqref{eq:CPclass} imposes constraints on the coupling terms in the bare Lagrangian. 

First, the minimal kinetic terms of bosons and fermions depend on both $\phi$ and its conjugate $\bar{\phi}$ in a symmetric manner. 
Recalling that the conjugate $\bar{\phi}$ is labeled by the conjugacy class $\overline{[g]}=[g^{-1}]$, the minimal kinetic terms are consistent with both the non-invertible selection rules and the CP transformation in general. 
The same conclusion applies to more generic kinetic terms if they are symmetric under the exchange $\phi \leftrightarrow \bar{\phi}$. 

Next, let us consider CP invariance of the interaction term of the form:
\begin{align}
    {\cal L}\supset \sum_s y_s\,(\phi_1...\phi_n)_{\mathbf{1},s} +{\mathrm{h.c.}},
\end{align}
where $y_s$ denotes complex coupling constants. 
Here and in what follows, we assume the existence of a group-like symmetry $G_f$ in the tree-level interaction of fields, 
under which each field is labeled by a certain conjugacy class of a group $G$. 
Furthermore, the label $\mathbf{1}$ and the index $s$ respectively represent an invariant singlet of $G_f$ and the independent singlets. 
In the basis $X^{\mathrm{CP}}=\mathbb{I}$, the interaction term transforms under CP as
\begin{align}
    y_s\,\phi_1...\phi_n 
    \xrightarrow{\mathrm{CP}} y_s^\ast\,\bar{\phi}_1...\bar{\phi}_n
    = y_s^\ast \overline{\phi_1...\phi_n},
\end{align}
where the reality of the Clebsch-Gordan coefficients is assumed in the equality. 
Hence, CP invariance of the theory requires the coupling $y_s$ to be real. 
It implies that mass matrices of quarks and leptons are also constrained to be real ones. 
Note that this discussion also holds at loop levels, if the CP symmetry associated with \eqref{eq:CPclass} is preserved at loop levels.

\section{Concrete examples}
\label{sec:ex}

In this section, we present explicit examples realizing the generalized CP in non-invertible selection rules arising from fusion algebras. 
In Sec.~\ref{sec:Z31}, we focus on the CP symmetry underlying certain non-invertible selection rules, and it can be combined with flavor symmetries in Sec.~\ref{sec:Z32}.

\subsection{CP}
\label{sec:Z31}

As a concrete example, let us consider $\mathbb{Z}_3$ gauging of $\mathbb{Z}_7$ symmetry. 
When the generator of $\mathbb{Z}_7$ is defined as $a$ with $a^7=e$, 
the $\mathbb{Z}_3$ automorphism $b$ of $\mathbb{Z}_7$ is described by
\begin{align}\label{eq:Z3-ZN}
      b^{-1}ab=a^2, \qquad b^3=e.
\end{align}
Note that the $\mathbb{Z}_3$ gauging of $\mathbb{Z}_N$ is possible only for $N=7,13,...$ (see, Ref.~\cite{Dong:2025jra}, for the derivation). 
Since the above action can be generalized to a generic element $a^k$, i.e., 
\begin{align}
    &b a^k b^{-1} = a^{4k}, \qquad 
    b^2 a^k b^{-2} = a^{16 k},
\end{align}
one can define the following classes:
\begin{align}   \label{eq:class-Z3} 
    &C^{(k)} := [a^k] = \{a^{2^l k} | l = 0, 2, 4\} = \{a^k, a^{2k}, a^{4k}=a^{-3k}\},
\end{align}
obeying 
\begin{align}\label{eq:product-Z3}
    C^{(k)}\cdot  C^{(\ell)} = C^{(k + \ell)} \oplus C^{(k + 4 \ell)} \oplus C^{(k + 16 \ell)} = C^{(k + \ell)} \oplus C^{(k - 3 \ell)} \oplus C^{(k + 2 \ell)}.
\end{align}
Given that the class $C^{(0)}$ includes the only identity, let us rewrite the classes: 
\begin{align}\label{eq:class-Z3-7}
  &C_1 := \{ e \},\quad
  C^1_3 := \{ a,a^2,a^4 \},\quad
  C^2_3 := \{ a^3,a^5,a^6 \},
\end{align}
whose multiplication rules are summarized in Table~\ref{tab:Z3gaugingZ7}. 
\begin{table}[H]
  \centering
  \caption{Multiplication rules of classes for $\mathbb{Z}_3$ gauging of $\mathbb{Z}_7$.}
    \label{tab:Z3gaugingZ7}
  \begin{tabular}{|c||c|c|c|}
  \hline
     & $C_1$ & $C^1_3$ & $C^2_3$ \\
     \hline\hline
     $C_1$&$C_1$ & $C^1_3$ & $C^2_3$ \\
     \hline
     $C^1_3$&$C^1_3$&$C^1_3+2C^2_3$   &$3C_1+C^1_3+C^2_3$  \\
     \hline
     $C^2_3$&$C^2_3$  &$3C_1+C^1_3+C^2_3$  &$2C^1_3+C^2_3$  \\
     \hline
  \end{tabular}
\end{table}

From the selection rules in Table~\ref{tab:Z3gaugingZ7}, one finds the existence of an $S_2\cong \mathbb{Z}_2$ permutation symmetry associated with
\begin{align}
C_3^1&\leftrightarrow C_3^2,
\end{align}
corresponding to the outer automorphism exchanging the group elements $a$ and $a^{-1}$. 
Consequently, when a field $\phi$ and its conjugate $\phi^\ast$ are respectively labeled by a class $C_3^1$ and $C_3^2$, this $\mathbb{Z}_2$ symmetry can be identified with the charge conjugation. 
In this concrete example, the CP transformation is defined in a basis with $X^{\mathrm{CP}}=\mathbb{I}$. 
Then, the simultaneous transformation of the charge conjugation and the parity transformation leads to the four-dimensional CP symmetry. 
It indicates that when all the Standard Model fields are labeled by the same conjugacy class, e.g. $C_3^1$, all the couplings are constrained to be real, as discussed in Sec.~\ref{sec:couplings}. 
Specifically, let us consider the CP-invariant Yukawa interaction:
\begin{align}
    {\cal L} = y_{ij} \bar{\psi}_{L,i} \psi_{R,j} H + \mathrm{h.c.},
\end{align}
where $i$ labels the fermion generations and $H$ denotes the Higgs fields or its complex conjugate. 
When $\bar{\psi}_L$, $\psi_R$ and $H$ are all labeled by the conjugacy class $C_1^1$, all Yukawa couplings $y_{ij}$ are allowed by the selection rules presented in Table~\ref{tab:Z3gaugingZ7}. 
The imposed CP symmetry then enforces these Yukawa couplings to be real. 
Note that the right-handed Majorana neutrinos are self conjugate under CP due to $\nu_{Ri}^C={\cal C}\overline{\nu}_{Ri}^T$ with ${\cal C}=i\gamma^0\gamma^2$ being the charge conjugation matrix. 
To realize the non-vanishing CKM and PMNS phases, the CP symmetry must be broken spontaneously. 
A concrete realization of spontaneous CP violation in this framework is discussed in detail in Sec.~\ref{sec:spontaneous}.

The existence of charge conjugation is not specific to the fusion rule presented in Table~\ref{tab:Z3gaugingZ7}. 
Indeed, the same $\mathbb{Z}_2$ symmetry appears in a variety of other fusion rules. 
For instance, the conjugacy classes arising from $S_3$ gauging of $\mathbb{Z}_3\times \mathbb{Z}_3$ obey a $\mathbb{Z}_3$ Tambara-Yamagami fusion rule, as summarized in Table~\ref{tab:Z3TY}, where the conjugacy classes are denoted by $\{C_1, C_1^1, C_1^2, C'_6 \}$ (for more details, see, Ref.~\cite{Dong:2025jra}). 
Remarkably, the corresponding selection rule exhibits an $S_2$ permutation symmetry under the exchange $C_1^1\leftrightarrow C_1^2$, which can be interpreted as a CP symmetry. 
Furthermore, the conjugacy classes of $A_4$ group also possess an analogous $S_2$ permutation symmetry, as explicitly shown in a later section. 
\begin{table}[H]
  \centering
  \caption{$\mathbb{Z}_3$ Tambara-Yamagami fusion rule.}
  \label{tab:Z3TY}
  \begin{tabular}{|c||c|c|c|c|}
  \hline
     &$C_1$ &$C^1_1$ &$C^2_1$ &$C'_6$ \\
     \hline
     \hline
     $C_1$&$C_1$ &$C^1_1$ &$C^2_1$ &$C'_6$ \\
     \hline
     $C^1_1$& $C^1_1$&$C^2_1$ & $C_1$& $C'_6$\\
     \hline
     $C^2_1$& $C^2_1$&$C_1$ &$C^1_1$ & $C'_6$\\
     \hline
     $C'_6$&$C'_6$ & $C'_6$&$C'_6$ &$C_1+C^1_1+C^2_1$ \\
     \hline
  \end{tabular}
\end{table}

\subsection{Flavor symmetry  and CP}
\label{sec:Z32}

So far, we have just focused on the CP transformation, but when there exists a flavor symmetry, it leads to the generalized CP transformation. 
For illustrative purposes, let us focus on $\mathbb{Z}_3$ gauging of $\mathbb{Z}_3\times \mathbb{Z}'_3$.

First, let us introduce the generators of $\mathbb{Z}_3\times \mathbb{Z}'_3$:
\begin{align}
    &a^3 = a'^3= e,\quad aa' = a'a,
\end{align}
with $e$ being the identity. By utilizing them, the $\mathbb{Z}_3\times \mathbb{Z}'_3$ elements are described by $a^\ell a'^m$ $(\ell,m=0,1,2)$ (mod $3$). 
Next, we introduce the $\mathbb{Z}_3$ element $b$ corresponding to the outer automorphism of the $\mathbb{Z}_3 \times \mathbb{Z}'_3$:
\begin{align}
    &b^3 = e,\quad bab^{-1} = a^{-1}a'^{-1},\quad ba'b^{-1} = a,\nonumber\\
    &b (a^\ell a'^m) b^{-1} = a^{-\ell + m} a'^{-\ell},\quad
    b^2 (a^\ell a'^m) b^{-2} = a^{-m} a'^{\ell - m}.    
\end{align}
It leads to the $\mathbb{Z}_3$ invariant classes:
  \begin{align}
\label{eq:Z3classofZNZN}
    &C^{(k,\ell)} := \{ b^n a^k a'^\ell b^{-n} | n=0,1,2  \} = 
    \{ a^ka'^\ell, a^{-k+\ell}a'^{-k}, a^{-\ell}a'^{k-\ell} \}.
  \end{align}
Hence, we have a total of five independent classes: 
  \begin{align}
    \begin{array}{ll}
    C_1 := \{ e \},&\\
    C^{s}_1 := \{ a^sa'^{-s} \}& (s = 1,2),\\
    C^{(k,0)}_3 := \{ a^k, a^{-k}a'^{-k}, a'^{k} \}\,\,& (k= 1,2),
  \end{array}
\end{align}
whose multiplication rules are summarized in Table \ref{tab:Z3gaugingZ3Z3}.
  \begin{table}[H]
    \centering
    \caption{Multiplication rules of classes for $\mathbb{Z}_3$ gauging of $\mathbb{Z}_3\times\mathbb{Z}'_3$.}
    \label{tab:Z3gaugingZ3Z3}
    \begin{tabular}{|c||c|c|c|c|c|}
    \hline
       & $C_1$ & $C^1_1$ & $C^2_1$ & $C^{(1,0)}_3$ & $C^{(2,0)}_3$ \\
       \hline
       \hline
       $C_1$& $C_1$ & $C^1_1$ & $C^2_1$ & $C^{(1,0)}_3$ & $C^{(2,0)}_3$ \\
       \hline
       $C^1_1$& $C^1_1$& $C^2_1$ & $C_1$ & $C^{(1,0)}_3$ & $C^{(2,0)}_3$ \\
       \hline
       $C^2_1$& $C^2_1$ &$C_1$  & $C^1_1$ & $C^{(1,0)}_3$ & $C^{(2,0)}_3$ \\
       \hline
       $C^{(1,0)}_3$& $C^{(1,0)}_3$ & $C^{(1,0)}_3$ & $C^{(1,0)}_3$ & $3C^{(2,0)}_3$ & $3C_1+3C^1_1+3C^2_1$ \\
       \hline
       $C^{(2,0)}_3$& $C^{(2,0)}_3$ & $C^{(2,0)}_3$ &$C^{(2,0)}_3$  & $3C_1+3C^1_1+3C^2_1$ & $3C^{(1,0)}_3$ \\
       \hline  
       \end{tabular}
  \end{table}
From Table~\ref{tab:Z3gaugingZ3Z3}, one finds that the selection rules of the conjugacy classes contain a $\mathbb{Z}_3$ flavor symmetry associated with
\begin{align}
   C_3^{(p,0)} \to \omega^p C_3^{(p,0)},
\end{align}
with $\omega = e^{2\pi i/3}$, originating from $a\rightarrow \omega a$ and $a'\rightarrow \omega a'$. In addition, there exists a $\mathbb{Z}_2$ symmetry:\footnote{There exists another $\mathbb{Z}_2$ symmetry $C_1^1\rightarrow C_1^2$, but we focus on Eq.~\eqref{eq:Z2_1}.}
\begin{align}
C_3^{(1,0)}&\leftrightarrow C_3^{(2,0)},    
\label{eq:Z2_1}
\end{align}
originating from $a \leftrightarrow a^{-1}$ and $a' \leftrightarrow a'^{-1}$. 
In a manner similar to the previous discussion of the canonical CP, the above $\mathbb{Z}_2$ symmetry can be identified with the charge conjugation when the fields $\phi$ and the conjugate $\phi^\ast$ are respectively labeled by the class $C_3^{(1,0)}$ and $C_3^{(2,0)}$. 
Hence, by combining it with the parity transformation, this $\mathbb{Z}_2$ symmetry can be identified by CP symmetry.

Remarkably, the $\mathbb{Z}_3$ symmetry identified with a flavor symmetry enhances the CP symmetry with $S_3\simeq \mathbb{Z}_3 \rtimes \mathbb{Z}_2$. 
This structure is understood as follows. 
Let us consider two scalar fields $\phi_1$ and $\phi_2$, which are labeled by
\begin{align}
    \phi =
    \begin{pmatrix}
        \phi_1\\
        \phi_2
    \end{pmatrix}
    .
\end{align}
They transform under the $\mathbb{Z}_3$ symmetry as
\begin{align}
    \begin{pmatrix}
        \phi_1\\
        \phi_2
    \end{pmatrix}
    \rightarrow
    \begin{pmatrix}
        \omega & 0 \\
        0 & \omega^2
    \end{pmatrix}
    \begin{pmatrix}
        \phi_1\\
        \phi_2
    \end{pmatrix}
    \equiv \rho
    \begin{pmatrix}
        \phi_1\\
        \phi_2
    \end{pmatrix}.
\end{align}
On the other hand, the CP transformation is described by
\begin{align}
    \begin{pmatrix}
        \phi_1\\
        \phi_2
    \end{pmatrix}
    \rightarrow
    \begin{pmatrix}
        \bar{\phi}_1\\
        \bar{\phi}_2
    \end{pmatrix}
    .
\end{align}
Hence, performing CP and flavor symmetry transformations, one finds
\begin{align}
    \phi_i \xrightarrow{\mathrm{CP}} \bar{\phi_i} \xrightarrow{\mathbb{Z}_3} \rho^{\ast}\phi_i \xrightarrow{\mathrm{CP}^{-1}} \rho^\ast\phi_i.
\end{align}
Although $\rho^\ast = \rho^{-1}$ is different from $\rho$, but it belongs to $\mathbb{Z}_3$ flavor symmetry. 
It indicates that the generalized CP transformation exists in this class of non-invertible selection rules. 
The full symmetry group is now isomorphic to $S_3\cong \mathbb{Z}_3\rtimes \mathbb{Z}_2^{\mathrm{CP}}$. 
When we enforce this symmetry to all the fields in a theory, all the couplings are constrained to be real, as discussed in Sec.~\ref{sec:couplings}. 

Before closing this section, we comment on the other selection rule having the generalized CP symmetry. 
Indeed, the existence of generalized CP is not special to the $\mathbb{Z}_3$ gauging of $\mathbb{Z}_3\times \mathbb{Z}'_3$. 
For instance, let us deal with the conjugacy classes of $A_4\cong(\mathbb{Z}_2 \times \mathbb{Z}'_2) \rtimes \mathbb{Z}_3$ group:
\begin{align}
    &C_1 := \{ e \},\\
    &C^{(1,0)}_3 := \{ a,a',aa' \},\\
    &C^1_4 := \{ b,ba,ba',baa' \},\\
    &C^2_4 := \{ b^2,b^2a,b^2a',b^2aa' \},
\end{align}
where $a$ and $a'$ respectively represent the generator of $\mathbb{Z}_2$ and $\mathbb{Z}_2'$, and $b$ denotes the generator of $\mathbb{Z}_3$, i.e., $a^2=a'^2=b^3=e$.  

From the multiplication rules of $A_4$ conjugacy class shown in Table~\ref{tab:Delta(12)}, 
we find two types of symmetries, as pointed out in Ref.~\cite{Dong:2025jra}:
\begin{enumerate}
    \item $\mathbb{Z}_2$ symmetry associated with $C_4^1 \leftrightarrow C_4^2$, i.e., the outer automorphism $b\leftrightarrow b^2$. 

    \item $\mathbb{Z}_3$ symmetry associated with $C_4^1 \rightarrow \omega C_4^1$ and $C_4^2\rightarrow \omega^2 C_4^2$, corresponding to $b \to \omega b$ with $\omega = e^{2\pi i/3}$.  
\end{enumerate}
When we identity the $\mathbb{Z}_2$ symmetry with the CP symmetry, the $\mathbb{Z}_3$ flavor symmetry and CP symmetry is enlarged to $S_3\cong \mathbb{Z}_3\rtimes \mathbb{Z}_2^{\mathrm{CP}}$, as in the case of $\mathbb{Z}_3$ gauging of $\mathbb{Z}_3\times \mathbb{Z}'_3$. 
In this work, we focus on Abelian flavor symmetries originating from the fusion algebras. However, fusion algebras can also give rise to non-Abelian flavor symmetries, as reported in Ref.~\cite{Dong:2025jra}. A comprehensive study of such possibilities is left for future work.

\begin{table}[H]
    \caption{Multiplication rules for conjugacy classes of $\Delta(12)$.}
    \label{tab:Delta(12)}
    \centering
    \begin{tabular}{|c||c|c|c|c|}
    \hline
    &$C_1$&$C^{(1,0)}_3$&$C^1_4$&$C^2_4$\\
    \hline\hline
    $C_1$&$C_1$&$C^{(1,0)}_3$&$C^1_4$&$C^2_4$\\
    \hline
    $C^{(1,0)}_3$&$C^{(1,0)}_3$&$3C_1+2C^{(1,0)}_3$&$3C^1_4$&$3C^2_4$\\
    \hline
    $C^1_4$&$C^1_4$&$3C^1_4$&$4C^2_4$&$4C_1+4C^{(1,0)}_3$\\
    \hline
    $C^2_4$&$C^2_4$&$3C^2_4$&$4C_1+4C^{(1,0)}_3$&$4C^1_4$\\
    \hline
    \end{tabular}
\end{table}

\section{Texture zeros with generalized CP}
\label{sec:textures}

In this section, we discuss the structure of Yukawa textures in the presence of CP symmetry. 
As mentioned above, one can construct a CP-invariant system when all the matter fields in a theory are appropriately assigned to the conjugacy class. Then, all entries of the three-point couplings, i.e., Yukawa couplings, are allowed by the fusion rules. 

For concreteness, let us consider $\mathbb{Z}_3$ gauging of $\mathbb{Z}_7$ discussed in Sec.~\ref{sec:Z31}. 
Since the conjugate of a field $\phi$ is labeled by the class $\overline{[g_i]}=[g_i^{-1}]$, the kinetic term of $\phi$ is always allowed, regardless of the specific assignments of the Standard Model matter fields under the fusion algebra. Hence, we focus on Yukawa interactions of quarks and leptons. 
In the quark sector, the Yukawa couplings are given by
\begin{align}
    {\cal L}_{\mathrm{quark}} = -(Y_u)_{ij}\bar{Q}_{L,i} H u_{R,j} - (Y_d)_{ij}\bar{Q}_{L,i} H^c d_{R,j},
\end{align}
with $H^c=i\sigma^2 H^\ast$. 
When all fields are assigned to the conjugacy class $C_1^1$ in the $\mathbb{Z}_3$ gauging of $\mathbb{Z}_7$, these Yukawa interactions are allowed by the fusion rules. 
Indeed, following the multiplication rule summarized in Table~\ref{tab:Z3gaugingZ7}, 
it turns out that all entries of Yukawa matrices are allowed in the CP-invariant system. 

For the lepton sector, we first consider the case in which neutrino masses are generated by the Weinberg operator: 
\begin{align}
    {\cal L}_W = -(Y_e)_{ij}\bar{L}_{L,i} H^c e_{R,j} - c_{ij}\frac{L_iHL_jH}{\Lambda}+\mathrm{h.c.},
\end{align}
with the cutoff scale $\Lambda$. 
In analogy with the quark sector, when we impose that charge leptons and Higgs field are assigned to to the conjugacy class $C_3^1$ under the $\mathbb{Z}_3$ gauging of $\mathbb{Z}_7$, all the couplings $\{(Y_e)_{ij},c_{ij}\}$ are allowed by the selection rules. 
On the other hand, the situation changes when we introduce right-handed neutrinos ($N_R$) to generate tiny active neutrino masses via the type-I seesaw mechanism~\cite{Minkowski:1977sc, Yanagida:1979as,Yanagida:1979gs,Gell-Mann:1979vob}. 
Since Majorana fermions are self conjugate under CP as mentioned before, both the right-handed neutrinos and their conjugates belong to the same class such as $C_3^1$ in the case of $\mathbb{Z}_3$ gauging of $\mathbb{Z}_7$. 
As a result, the Majorana mass term $M_{ij}\bar{N}_{R,i}^cN_{R,j}$ is forbidden by the corresponding selection rule. 
However, when we introduce a Standard Model singlet scalar field $\Phi$, the following interactions are allowed
\begin{align}
    {\cal L}_{\nu}= -(Y_e)_{ij}\bar{L}_{L,i} H^c e_{R,j} -(Y_\nu)_{ij}\bar{L}_{L,i} H N_{R,j} - \frac{1}{2}(M_N)_{ij} \Phi \bar{N}^c_{R,i} N_{R,j} + \mathrm{h.c.}.
\end{align}
Here, all the fields are again assigned to the conjugacy class $C_3^1$.

In the following, we discuss the Yukawa textures of quarks and leptons in a framework where the neutrino masses are generated via the type-I seesaw mechanism with three right-handed neutrinos. 
In addition to the CP symmetry realized through $\mathbb{Z}_3$ gauging of $\mathbb{Z}_7$, 
we impose further selection rules, in particular, $\tilde{\mathbb{Z}}_5^{(1)}\times \tilde{\mathbb{Z}}_5^{(2)}$.\footnote{Here and in what follows, we call the $\mathbb{Z}_2$ gauging of $\mathbb{Z}_5$ as $\tilde{\mathbb{Z}}_5$.} 
Note that for $\mathbb{Z}_2$ gauging of $\mathbb{Z}_5$ (denoted by $\tilde{\mathbb{Z}}_5$), there exist three conjugacy classes:
\begin{align}
    [g^0] &= \{g^0\}\,,
    \nonumber\\
    [g^1] &= \{ g^1, g^4\}\,,
    \nonumber\\
    [g^2] &= \{ g^2, g^3\}\,,
\end{align}
where $g^i$ with $i=1,2,3,4$ corresponds to the $\mathbb{Z}_5$ generator. 
They obey the following fusion rule:
\begin{align}
    [g^0]\cdot[g^0] &= [g^0]\,,
    \nonumber\\
    [g^0]\cdot[g^m] &= [g^m]\cdot[g^0] = [g^i]\,,
    \nonumber\\
    [g^1]\cdot[g^1] &= [g^0] \oplus [g^2]\,,
    \nonumber\\
    [g^1]\cdot[g^2] &= [g^2]\cdot[g^1] = [g^1] \oplus [g^2]\,,
    \nonumber\\
    [g^2]\cdot[g^2] &= [g^0] \oplus [g^1]\,,
\end{align}
with $m=1,2$.

Let us take the assignment of matter fields as follows. 
For $\tilde{\mathbb{Z}}_5^{(1)}$, we choose 
\begin{align}
    Q_i&:([g^0],[g^1],[g^2]),\qquad U_i:([g^0],[g^1],[g^2]),\qquad
    D_i:([g^0],[g^1],[g^2]),
    \nonumber\\
    L_i&:([g^0],[g^1],[g^2]),\qquad E_i:([g^1],[g^0],[g^1]),\qquad
    N_i: ([g^0],[g^1],[g^2])
    \nonumber\\
    H &: [g^1],\qquad\qquad \qquad
    \Phi : [g^0],
\end{align}
which leads to
\begin{align}
Y_{u}&=
\begin{pmatrix}
0 & * & 0 \\
* & 0 & *  \\
0 & * & *
\end{pmatrix}
,\qquad Y_{d}=
\begin{pmatrix}
0 & * & 0 \\
* & 0 & *  \\
0 & * & *
\end{pmatrix},
\nonumber\\
    Y_{e}&= 
    \begin{pmatrix}
* & 0 & * \\
0 & * & 0  \\
* & 0 & *
    \end{pmatrix},
    \qquad
    Y_{\nu}= 
    \begin{pmatrix}
0 & * & 0 \\
* & 0 & *  \\
0 & * & *
    \end{pmatrix},
    \qquad
    M_N= 
    \begin{pmatrix}
* & 0 & 0 \\
0 & * & 0  \\
0 & 0 & *
    \end{pmatrix},    
\end{align}
where $\ast$ denotes the non-zero entries. 

For $\tilde{\mathbb{Z}}_5^{(2)}$, we consider the following assignments of matter fields:
\begin{align}
    Q_i&:([g^2],[g^0],[g^0]),\qquad U_i:([g^1],[g^2],[g^1]),\qquad
    D_i:([g^1],[g^1],[g^1]),
    \nonumber\\
    L_i&:([g^2],[g^1],[g^1]),\qquad E_i:([g^1],[g^2],[g^0]),\qquad N_i:([g^0],[g^2],[g^2]),
    \nonumber\\
    H &: [g^1],\qquad\qquad \qquad
    \Phi : [g^0],
\end{align}
which leads to
\begin{align}
Y_{u}&=
\begin{pmatrix}
* & * & * \\
* & 0 & *  \\
* & 0 & *
\end{pmatrix}
,\qquad Y_{d}=
\begin{pmatrix}
* & * & * \\
* & * & *  \\
* & * & *
\end{pmatrix},
\nonumber\\
    Y_{e}&= 
    \begin{pmatrix}
* & * & 0 \\
0 & * & *  \\
0 & * & *
    \end{pmatrix},
    \qquad
    Y_{\nu}= 
    \begin{pmatrix}
0 & * & * \\
* & * & *  \\
* & * & *
    \end{pmatrix},
    \qquad
    M_N= 
    \begin{pmatrix}
* & 0 & 0 \\
0 & * & *  \\
0 & * & *
    \end{pmatrix},    
\end{align}

By combining the selection rules of $\tilde{\mathbb{Z}}_5^{(1)}\times \tilde{\mathbb{Z}}_5^{(2)}$, we arrive at the following Yukawa textures:
\begin{align}
Y_{u}&=
\begin{pmatrix}
0 & * & 0 \\
* & 0 & *  \\
0 & 0 & *
\end{pmatrix}
,\qquad Y_{d}=
\begin{pmatrix}
0 & * & 0 \\
* & 0 & *  \\
0 & * & *
\end{pmatrix},
\nonumber\\
    Y_{e}&= 
    \begin{pmatrix}
* & 0 & 0 \\
0 & * & 0  \\
0 & 0 & *
    \end{pmatrix},
    \qquad
    Y_{\nu}= 
    \begin{pmatrix}
0 & * & 0 \\
* & 0 & *  \\
0 & * & *
    \end{pmatrix},
    \qquad
    M_N= 
    \begin{pmatrix}
* & 0 & 0 \\
0 & * & 0  \\
0 & 0 & *
    \end{pmatrix},   
\end{align}
where the explicit values of $Y_u$ and $Y_d$ are written in, e.g., eq.(6.19) of Ref.~\cite{Kobayashi:2025znw}, and the lepton sector was discussed in Ref.~\cite{Kaneta:2016gbq}.

\section{Spontaneous CP violation}
\label{sec:spontaneous}


As a simplest example, we consider a model based on $\mathbb{Z}_3$ gauging of $\mathbb{Z}_7$. 
As discussed in Sec.~\ref{sec:Z31}, the fusion algebra admits an $S_2$ permutation symmetry associated with $C_1^1\leftrightarrow C_1^2$, and it can be identified with the charge conjugation when a field and its conjugate are respectively labeled by the conjugacy classes $C_1^1$ and $C_1^2$. 
As an illustration, let us consider a Standard Model singlet scalar field $\Phi$ which is labeled by the conjugacy class $C_1^1$. 
Then, the CP-invariant scalar potential at tree level is given by
\begin{align}
\label{eq:V_general}
    V &= m^2 \Phi \Phi^\ast + \xi_1 \Phi\Phi^\ast(\Phi + \Phi^{\ast}) + \xi_2 (\Phi^3 + \Phi^{\ast3}) + \lambda_1 (\Phi \Phi^\ast)^2 
    \nonumber\\
    &+ \lambda_2 \Phi \Phi^\ast(\Phi^2 + \Phi^{\ast2})+\lambda_3 (\Phi^4+\Phi^{\ast 4}).
\end{align}
Depending on the signs and magnitudes of the parameters, the vacuum configuration may spontaneously break the CP symmetry. 
Since obtaining analytic solutions for the vacuum expectation value (vev) of $\Phi$ is in general difficult for generic parameters, we first present analytic CP-breaking solutions in a limited region of parameter space, and subsequently perform a numerical analysis for the generic case. 
\begin{itemize}
    \item $\lambda_2=\lambda_3=0$

When we focus on $\lambda_2=\lambda_3=0$, the potential of the angular direction of $\Phi$, i.e., $\theta:=\arg(\Phi)$, is given by
\begin{align}
    V \supset 2r^3\left(\xi \cos 3\theta + \xi' \cos \theta\right),
\end{align}
with $r=|\Phi|$. 
The vev of $\theta$ is then determined by
\begin{align}
    \theta = 
    \left\{
    \begin{array}{ll}
         \pi & (\xi_1,\xi_2>0)\\
         \frac{1}{2}\cos^{-1}\left(\frac{-3\xi - \xi'}{6\xi} \right) & (\xi_1>0,\xi_2<0)\\
         \pi + \frac{1}{2}\cos^{-1}\left(\frac{-3\xi - \xi'}{6\xi} \right)& (\xi_1<0,\xi_2>0)\\
         0 & (\xi_1,\xi_2<0)\\
    \end{array}
    \right.
    .
\end{align}
It indicates that the signs of $\xi$ and $\xi'$ determine whether the CP symmetry is spontaneously broken at the minimum. 
The vev for the radial direction of $\Phi$ is found as
\begin{align}
    r = \frac{(3 \xi_2-\xi_1)^{3/2} + \sqrt{(3 \xi_2 - \xi_1)^3 - 
  24 m^2 \lambda_1 \xi_2}}{4  \lambda_1\sqrt{3\xi_2}},
\end{align}
where we consider the case $\xi_1>0$ and $\xi_2<0$. 
By performing the numerical analysis with the parameters:
\begin{align}
    m^2 = -1,\quad  \xi_1=-1,\quad \xi_2=1,\quad \lambda_1=5,
\end{align}
one finds that the CP breaking minimum 
\begin{align}
\langle r\rangle \simeq 0.62,\qquad
\langle \theta \rangle \simeq 0.96
\end{align}
is a global minimum, and the Hessian matrix at this minimum is positive definite. 

    \item $\xi_1=\xi_2=0$

In this case, the potential of the angular direction of $\Phi$ is given by
\begin{align}
    V \supset 2r^4\left(\lambda_2 \cos 2\theta + \lambda_3 \cos 4\theta\right),
\end{align}
with $r=|\Phi|$. Similarly, the signs of $\lambda_{1,2}$ determine the nature of CP-breaking at the minimum. 
When we take $\lambda_1>0$ and $\lambda_2<0$, the vev of $\theta$ is determined by
\begin{align}
    \cos\theta = 
    \pm \sqrt{\frac{-\lambda_2 + 4 \lambda_3}{8\lambda_3}}
    .
\end{align}
In addition, the vev of the radial direction is fixed at
\begin{align}
    r=\sqrt{\frac{2 m^2 \lambda_3}{\lambda_2^2 - 
 4 (\lambda_1 - 2 \lambda_3) \lambda_3}}.
\end{align}
We numerically check the vacuum structure of $\Phi$. 
By taking
\begin{align}
    m^2 = -\frac{1}{2},\quad  \lambda_1=5,\quad \lambda_2=-3,\quad
    \lambda_3=1,
\end{align}
we find that the CP breaking minimum 
\begin{align}
\langle r\rangle \simeq 0.58,\qquad
\langle \theta \rangle \simeq 0.36
\end{align}
is a global minimum, and the Hessian matrix at this minimum is positive definite. 

\end{itemize}

So far, we have focused on restricted cases. However, our results indicate that the CP symmetry can be spontaneously broken, at least for small values of $\lambda_{2,3}$ or $\xi_{1,2}$. 
To understand the vacuum structure of CP-breaking vacua in the presence of non-vanishing $\lambda_{2,3}$ or $\xi_{1,2}$, we perform a numerical analysis of the scalar potential. 
For the generic scalar potential \eqref{eq:V_general}, the minimum along the radial direction of $\Phi$, i.e., $r=|\Phi|$, depends on the angular direction $\theta$. 
Extremizing the potential with respect to $r$, one finds
\begin{align}
    r &= \frac{\xi_1 \sin(\theta) - 3 \xi_2 \sin(3 \theta)}{
2 \lambda_2 \sin(2 \theta) + 4 \lambda_3 \sin(4 \theta)}.
\end{align}
As an explicit example, we take
\begin{align}
    m^2 = -1,\quad \xi_1=-2,\quad \xi_2=1,\quad \lambda_1=6,\quad \lambda_2=\lambda_3=\frac{1}{2},
\end{align}
leading to a CP-breaking global minimum located at
\begin{align}
\langle r\rangle \simeq 0.79,\qquad
\langle \theta \rangle \simeq 0.89.
\end{align}
The Hessian matrix at this minimum is positive definite.

Let us comment on one possible mechanism by which the CP-breaking effect mediates to the Standard Model sector. 
We have addressed the spontaneous CP violation triggered by the scalar field $\Phi$, which is labeled by the conjugacy class $C_1^1$ in the $\mathbb{Z}_3$ gauging of $\mathbb{Z}_7$. 
However, in general $\Phi$ may carry additional labels under other fusion algebras beyond those having CP. 
Following the discussion in Sec.~\ref{sec:textures}, let us suppose that $\Phi$ is also labeled by the conjugacy class $[g^0]$ under the $\mathbb{Z}_2$ gauging of $\mathbb{Z}_5$. The scalar potential of $\Phi$ remains unchanged and is still given by Eq.~\eqref{eq:V_general}.

Since the field $\Phi$ is supposed to be the Standard Model singlet field, it can couple with the Standard Model fermions through
\begin{align}
    {\cal L} &= -(Y_u^{'})_{ij}\frac{\Phi}{\Lambda}\bar{Q}_{L,i} H u_{R,j} - (Y_d^{'})_{ij}\frac{\Phi}{\Lambda}\bar{Q}_{L,i} H^c d_{R,j}
 -(Y_e^{'})_{ij}\frac{\Phi}{\Lambda}\bar{L}_{L,i} H^c e_{R,j} -(Y_\nu^{'})_{ij}\frac{\Phi}{\Lambda}\bar{L}_{L,i} H N_{R,j} 
 \nonumber\\&+ \mathrm{h.c.},
\end{align}
where $Y_f^{'}$ with $f=\{u,d,e,\nu\}$ are real values, and $\Lambda$ denote the cutoff scale. 
Hence, below the mass scale of $\Phi$, the effective Yukawa couplings take the form:
\begin{align}
\label{eq:Yeff}
    Y_f^{\mathrm{(eff)}}=Y_f + Y_f^{'}\frac{\Phi}{\Lambda},
\end{align}
which become complex as a consequence of the spontaneous CP violation. 
Since not all complex phases can be removed by field redefinitions, the effective Yukawa couplings contain physical CP-violating phases.

As an illustrative example, let us consider $\tilde{\mathbb{Z}}_5^{(1)}\times \tilde{\mathbb{Z}}_5^{(2)}$ as well. 
When the left-handed quarks, right-handed quarks and Higgs fields are labeled by
\begin{align}
    \mathrm{Left}&: ([g^1], [g^1], [g^2]),\qquad
    \mathrm{Right}: ([g^1], [g^2], [g^2]),\qquad \mathrm{Higgs}:[g^1]\quad \mathrm{for\,\,\tilde{\mathbb{Z}}_5^{(1)}},
    \nonumber\\
    \mathrm{Left}&: ([g^1], [g^2], [g^2]),\qquad
    \mathrm{Right}: ([g^2], [g^1], [g^2]),\qquad \mathrm{Higgs}:[g^1]\quad \mathrm{for\,\,\tilde{\mathbb{Z}}_5^{(2)}},
\end{align}
the up- and down-type Yukawa matrices take the form:
\begin{align}
\label{eq:3zero}
    Y_{u,d}=
        \begin{pmatrix}
0 & 0 & * \\
0 & * & *  \\
* & * & *
    \end{pmatrix},
\end{align}
where $\ast$ denotes non-vanishing real numbers. 
In addition, let us introduce the Standard Model singlet scalar field $\Phi$ labeled by the identity class:
\begin{align}
    \Phi:[g^0]\quad \mathrm{for\,\,\tilde{\mathbb{Z}}_5^{(1,2)}},
\end{align}
whose scalar potential is still identical to Eq.~\eqref{eq:V_general}. 
Hence, one can achieve the spontaneous CP violation for appropriate choices of parameters. 
Since $\Phi$ is labeled by the identity class, 
the textures of higher-dimensional operators
\begin{align}
    -(Y_u^{'})_{ij}\frac{\Phi}{\Lambda}\bar{Q}_{L,i} H u_{R,j} - (Y_d^{'})_{ij}\frac{\Phi}{\Lambda}\bar{Q}_{L,i} H^c d_{R,j}+\mathrm{h.c.}
\end{align}
are the same with the renormalizable Yukawa couplings~\eqref{eq:3zero}, i.e., 
\begin{align}
    Y^{'}_{u,d}=
        \begin{pmatrix}
0 & 0 & * \\
0 & * & *  \\
* & * & *
    \end{pmatrix}
    ,
\end{align}
where it is notable that $\ast$ again denotes non-vanishing real numbers, but its value can be different from Eq.~\eqref{eq:3zero}. 
Hence, $Y^{'}_{u,d}$ also exhibit the same three-zero texture. 
We focus on the specific Yukawa texture, but this discussion is also applicable to other textures if we take the class of $\Phi$ as the identity class, as in the example of Sec.~\ref{sec:textures}. 
As a result, even after the spontaneous CP violation induced by $\Phi$, the effective Yukawa couplings~\eqref{eq:Yeff} still have the same Yukawa textures as $Y_f$.\footnote{If the class of $\Phi$ is not labeled by the identity class, the effective Yukawa couplings will have the different structure from $Y_f$. 
}

\section{Conclusions}
\label{sec:con}

A new class of selection rules based on fusion algebras has recently received attention in flavor physics. 
In particular, Yukawa textures arising from non-invertible selection rules cannot be realized within conventional group-based symmetries, as explicitly demonstrated in e.g. Refs.~\cite{Kobayashi:2024cvp}. 
Despite this progress, the relation between the CP transformation and these non-invertible selection rules has not been fully clarified yet. 
It motivated us to construct a CP-invariant framework in which fields are labeled by specific basis elements of a fusion algebra endowed with a $\mathbb{Z}_2$ symmetry. 

In this paper, we focused on a framework in which fields are labeled by conjugacy classes of a finite group rather than by its irreducible representations. 
The resulting selection rules for fields cannot be captured by those arising from group-based symmetries. 
However, the selection rules associated with certain conjugacy classes contain a $\mathbb{Z}_2$ symmetry related to the outer automorphism of the group. 
We identified this $\mathbb{Z}_2$ symmetry underlying the fusion algebra with the charge conjugation. Together with the parity transformation, one can define a CP symmetry. 
Furthermore, certain fusion algebras admit group-based flavor symmetries. 
By combining group-based flavor symmetries and CP symmetry, one can define the generalized CP even within the context of quantum field theories where fields are labeled by the conjugacy classes. 
In such a CP-invariant setup, the couplings are then constrained to be real, in analogy with models based on conventional group-based flavor symmetries. 
We also proposed a mechanism for spontaneous CP violation triggered by a Standard Model singlet field $\Phi$, which mediates the CP-breaking effects to the Standard Model sector.

When additional selection rules based on a fusion algebra are imposed in addition to CP, one can restrict Yukawa matrices to specific textures. 
It indicates that the texture zero approach can be consistently formulated within a CP-invariant framework. 
As a result, one can derive testable predictions for the flavor structure of quarks and leptons, in analogy with conventional flavor models. 
Although there are several directions to pursue, we focused on one phenomenological implication of non-invertible selection rules with CP, i.e., their relevance to the texture-zero approach. 
In particular, we studied the three-zero textures \eqref{eq:3zero} realized by the non-invertible selection rules arising from $\mathbb{Z}_2$ gauging of $\mathbb{Z}_5$.  It turned out that these textures still hold even after the spontaneous CP violation. 
This example provides a concrete phenomenological application of non-invertible selection rules in the presence of CP. 
It is fascinating to apply this new class of selection rules to various aspects of flavor physics, which is left for future work. 

We finally comment on the strong CP problem. 
When we impose the CP symmetry in a theory, the strong CP phase 
\begin{align}
    \bar{\theta} = \theta_0 + \arg\det(M_uM_d)
\end{align}
vanishes at the ultraviolet scale. Here, $\theta_0$ denotes the bare QCD CP phase. 
Although the non-vanishing CKM and PMNS phases are generated through the spontaneous CP violation as seen in this paper, generic Yukawa textures induce the strong CP phase $\arg\det(M_uM_d)$. 
So far, various Yukawa textures that avoid inducing the strong CP phase have been studied in the context of toroidal orbifold compactification~\cite{Liang:2024wbb}, the modular flavor models~\cite{Feruglio:2023uof,Petcov:2024vph,Penedo:2024gtb}, and models with non-invertible selection rules~\cite{Liang:2025dkm,Kobayashi:2025thd,Kobayashi:2025rpx}. It would also be interesting to derive Yukawa textures that reconcile the absence of the strong CP phase with the spontaneous CP violation, while the weak CP phase is successfully generated. The detailed discussion in this direction will be reported elsewhere.

\acknowledgments

We would like to thank Ye-Ling Zhou for his contributions in the early stages of this work. 
This work was supported by JSPS KAKENHI Grant Numbers JP23K03375 (T.K.), JP25H01539 (H.O.).

\bibliography{references}{}
\bibliographystyle{JHEP}

\end{document}